\documentclass[3p]{elsarticle}
\usepackage{amsmath}
\usepackage{amssymb}
\usepackage{mathrsfs}
\usepackage{graphicx}
 \usepackage{mathptmx}
  \usepackage[T1]{fontenc}
\usepackage[latin9]{inputenc}
\usepackage{xcolor}
\usepackage{pdfcolmk}
\usepackage{textcomp}
\usepackage{amsthm}
\usepackage{stackrel}
\usepackage{esint}
\usepackage{float}

\usepackage{lineno,hyperref}
\modulolinenumbers[5]

\journal{}

\newtheorem{thm}{Theorem}
\newtheorem{lem}[thm]{Lemma}
\newdefinition{rmk}{Remark}
\newproof{pf}{Proof}
\newproof{pot}{Proof of Theorem \ref{thm2}}

\begin{document}

\begin{frontmatter}

\title{The Novel Approach to the Closed-Form Average Bit Error Rate Calculation for the Nakagami-m Fading Channel\tnoteref{t1}}
\tnotetext[t1]{\copyright 2022. This manuscript version is made available under the CC-BY-NC-ND 4.0 (\href{license https://creativecommons.org/licenses/by-nc-nd/4.0/}{license https://creativecommons.org/licenses/by-nc-nd/4.0/})}



\author[mymainaddress]{Aleksey~S.~Gvozdarev}
\cortext[mycorrespondingauthor]{Corresponding author}
\ead{asg.rus@gmail.com}

\address[mymainaddress]{P.~G.~Demidov Yaroslavl State University, Yaroslavl, 150003, Russia}

\begin{abstract}
The research presents a procedure of the closed-form average bit error rate evaluation for wireless communication systems in the presence of multipath fading. A generalization of the classical moment generating function is applied, and a connection between the Hankel-type contour-integral representation of the Marcum Q-function and Gauss Q-function is obtained and applied to the analytic derivation of the quadrature amplitude modulated signal average bit error rate in the presence of Nakagami-m fading. The correctness of the obtained solution was verified, and its computational efficiency (in terms of accuracy and time gain), compared with the prevailing approximation, was demonstrated. The proposed methodology can be extended to a wide variety of composite channels.
\end{abstract}

\begin{keyword}
Gaussian Q-function \sep contour integral \sep closed-form \sep average bit error rate \sep Nakagami-m \sep multipath fading channel
\end{keyword}

\end{frontmatter}


\section{Introduction}

Modern wireless communication systems (for instance, 5th generation standard (5G) \cite{Kim20}) and services \cite{Van20}, like ultra-reliable low latency communications (URLLC) and enhanced mobile broadband (eMBB), place heavy demands on the spectrum resource allocation, which can be addressed by employing such approaches as carrier aggregation and ultrawideband signal application. At the same time, the usage of those methodologies raises the profiles of accounting wireless channel fading effects and their impact on communication link quality and reliability, which is usually described in terms of bit error rate (BER) and its averaged (over the fluctuating instantaneous signal-to-noise ratio (SNR)) version (ABER) \cite{Sim05,Pro07}.

A classical approach to considering the fading effects resides in the application of various multipath propagation channel models \cite{Sim05,Pro07,Pep11} among which one can mention two families: the classic models (like Rayleigh, Rician, Hoyt, Weibull, etc.) and composite ones (for example, $\kappa-\mu$, $\alpha-\mu$, $\eta-\mu$, generalized-gamma, etc.). Nevertheless, the composite models generalize the classical incorporating them as the specific limiting cases and are robust to the change in fading conditions they are given in terms of very complex mathematical description, which in many cases prevents from obtaining closed-form analytical expressions for link quality; hence, various simplifying assumptions and approximations are used.

Thus, in most of these applications, a balance between analytical description simplicity and correspondence with the real-life channel measurements is required. One of the possible solutions is the widely renowned Nakagami-m fading model \cite{Nak60}. Although proposed in the middle of the 20th century, it is still extensively used in various applications \cite{Que17, Sil19, Ahm19} delivering high adequacy with measurements. Its main advantage is the combination of the relative simplicity of the mathematical treatment \cite{Pep11} and the ability to cover a wide range of fading scenarios: it successfully handles light fading (turning into the Rician model in the extreme case \cite{Sim05}), classical Rayleigh fading (being a specific subcase of Nakagami-m \cite{Sim05}) and heavier than Rayleigh fading (the so-called hyper-Rayleigh \cite{Rao15}). The last one has drawn specific attention in the last few years in the light of the ad hoc communication systems (see, for instance, \cite{Gar19, AlJ21}).

The main problem with the direct implementation of that assumed model is that the derivation of the ABER expression requires the solution of integrals involving the Gaussian Q-function and its square. Despite the fact that the first one is widely addressed in the literature \cite{Sim05, Won04, Sam12, Shi13, Sur07, Que10}, the second has been reported only several times \cite{Sim05, Mal08, Sou12, Sou13}. The approach of computing integrals with the squared Gaussian Q-function (proposed initially in \cite{Mal08}) exhibits a specific flaw of applicability only to the cases of integer and half-integer Nakagami fading parameter $m$, hence doesn't convey the important case of hyper-Rayleigh fading ($m<1$). Later, it was elaborated in \cite{Sou12, Sou13} and applied to the Generalized K model (and its extended version), delivering the closed-form expressions in terms of the multivariate Fox H-function and Meijer G-function that are not common in modern computer algebra systems and though can be treated numerically by multiple series representation, cause problems with their truncation, error estimation and practical application.
On the other hand, numerous researches proposed a wide variety of numerically efficient Gaussian Q-function approximations, among which one can mention piecewise linear approximations (proposed recently in \cite{Cai20, Cai21}) and well-established exponential-type approximations (see, for instance, \cite{Chi03, Sad17, Los09, Ola12}).
Although their application can help derive easy to implement expressions, their accuracy depends heavily upon channel and signal models. Furthermore, for propagation conditions with heavy fading combined with higher-order modulations (for example, for 4096-level quadrature amplitude modulation (4k-QAM) \cite{Sch19, Den20}), the approximation quality sufficiently degrades, thus, leaving with under/overestimation of the overall system performance quality.
Thus, closed-form analytical representation (valid for arbitrary system parameters) or its computationally efficient approximation of the ABER for the Nakagami-m fading model is still a highly relevant problem.

At the same time, in detection theory integrals involving similar functions were successfully treated via the classical moment generating function (MGF) approach \cite{Sim98a, Pro07, Sim05}, which relates the arising special functions (namely, Marcum Q-function and Toronto function) with the closed-loop integral that is further applied to derive, for example, the probability of detection \cite{Her11a,Chu18,Gvo21a, Gvo22}, or bit error \cite{Pro07,Odr09,Sim98a}) as the sum (finite or infinite) of residues inside this loop (see \cite{Tel00,Her11b,Gvo21a,Chu18}).
It was demonstrated that such an approach is valid only for specific cases: most of the system and the channel parameters and their combinations must be integer-value. Hence, the feasible applications of the derived expressions are heavily confined.
A possible remedy to this problem was proposed in \cite{Gvo17} and was successfully applied for the extension of the classical contour-integral MGF approach.

The correspondence proposes an approach for evaluating the integrals involving the Gauss Q-function and its square. It relies upon the generalized MGF methodology and the relation between some special functions arising in many problems of wireless communications. By deriving such a connection and applying their contour-integral representations, a novel form for closed-form ABER and its approximate version for Nakagami-m fading channel is evaluated. Lastly, a numeric simulation was performed to demonstrate their computational efficiency (in terms of computational error and time).

The research is structured as follows: Section 2 gives a brief review of some preliminary aspects of ABER definition and its approximation widely used in the literature; Section 3 discusses the derived expressions for the integrals involving the Gauss Q-function and its square: general integral form, stated in Lemma 1, and applied to the Nakagami-m fading model, stated in Lemma 2 and Lemma 3, and the final result of the ABER for the assumed model (closed-form and approximation) stated in Theorem 4; Section 4 demonstrates the correctness and efficiency of the derived representations; Section 5 gives final conclusions.

\section{Preliminaries}

Let us assume that a microwave signal propagates through a wireless random fading channel with the probability
density function of the instantaneous SNR given by $f_{\gamma }\left(\gamma \right)$. Thus, the ABER can be defined as

\begin{equation}
\mathrm{\overline{BER}}=\int_{0}^{\infty}\mathrm{BER}\left(\gamma \right)f_{\gamma }\left(\gamma \right)d\gamma,\label{eq:ABER}
\end{equation}
with $\mathrm{BER}\left(\gamma \right)$ being the instantaneous BER, calculated for most of the modern modulation types in terms of Gauss Q-function or squared Gauss Q-function,
which in many cases prevents closed-form analytic evaluation
of (\ref{eq:ABER}), for instance, in the case of M-ary quadrature amplitude
modulation (see, for instance, \cite{Pro07,Sim05}):
\begin{equation}\label{eq:BERth}
\mathrm{BER}_{th}\left(\gamma \right)=4c_{0}Q\left(\sqrt{2c_{1}\gamma }\right)-4c_{0}^{2}Q^{2}\left(\sqrt{2c_{1}\gamma }\right),
\end{equation}
where $c_{0}=\frac{(\sqrt{M}-1)}{\sqrt{M}\log_2(M)}$ and $c_{1}=\frac{3\log_2(M)}{2(M-1)}$.
To cope with this problem, various approximations were proposed, hence
yielding closed-form results (see \cite{Sim05}, section~5.1.4).
According to the prevailing one
\begin{equation}
\mathrm{BER}_{ap}\left(\gamma \right)=a_{1}\sum_{j=1}^{a_{3}}Q\left(\sqrt{a_{2}(j)\gamma }\right)\label{eq:BERap}
\end{equation}
with the set of coefficients $\left\{ a_{1},a_{2}(j),a_{3}\right\} $
explicitly defined for specific modulation (see \cite{Lu99,Sim05}),
for example in case of M-QAM: $\left\{ a_{1},a_{2}(j),a_{3}\right\} =\left\{ 4c_{0},c_{1}(2j-1)^{2},\sqrt{M}/2\right\}$.

Although (\ref{eq:BERap}) makes it possible to avoid squared Gauss
Q-function the result is satisfactory only in high SNR regime and
low-dimensional modulation. The situation worsens when (\ref{eq:BERap})
is being averaged in (\ref{eq:ABER}) and the discrepancy between
(\ref{eq:BERap}) and (\ref{eq:BERth}) increases. Moreover, for multipath
fading channels, the SNR factor comes as a multiplier of fading parameter.
Hence for deep fading, high-dimensional modulation and small SNR regime
(which is of primary interest for modern wireless communication systems,
like 5G), the approximation results are usually unsatisfactory. Consequently,
it is important to have at hand a possible closed-form solution rather
than its approximation.

\section{Derived results}

As it was stated, the problem of evaluating \eqref{eq:ABER}, with BER being given by \eqref{eq:BERth} or \eqref{eq:BERap}, can be tackled with the help of the extended MGF approach. The key point is to find a relation between the arising Gauss Q-function and its suitable contour-integral representation, given by the following Lemma.

\begin{lem}
The Gauss Q-function can be represented as an integral in the complex
domain
\begin{equation}
Q\left(z\right)=\frac{1}{2}\frac{1}{2\pi i}e^{-\frac{z^{2}}{2}}\int_{-\infty}^{\left(0_{+}\right)}\frac{e^{\frac{z^{2}p}{2}}}{\sqrt{p}\left(1-p\right)}\mathrm{d}p\label{eq:Lem1}
\end{equation}
 with the integration Hankel-type integration contour.
\end{lem}
\begin{pf}
The representation \eqref{eq:Lem1} can be proved establishing the connection between the Gauss Q-function and the generalized
Marcum Q-function, and then by applying the contour-integral representation of the latter (see \cite{Gvo17}):
\begin{eqnarray}
\hspace{-10pt}Q\left(z\right)=\frac{1}{2}Q_{\frac{1}{2}}\left(0,z\right)\!\!\!\!\!\!&=&\!\!\!\!\!\!\frac{1}{2}\frac{1}{2\pi i}e^{-\frac{z^{2}}{2}}\!\!\int_{-\infty}^{\left(0_{+}\right)}\!\!\!\!\frac{e^{\frac{z^{2}p}{2}}}{\sqrt{p}\left(1-p\right)}\mathrm{d}p\nonumber\\
&=&\!\!\!\!\frac{1}{2}\left\{ \frac{1}{2\pi i}e^{-\frac{z^{2}}{2}}\right.\stackrel{\mathrm{Q}_{1}\left(\eta,z\right)}{\overbrace{\underset{C_{\eta }\left\backslash \left\{ -\eta \right\} \right.}{\ointctrclockwise}\frac{e^{\frac{z^{2}p}{2}}}{\sqrt{p}\left(1-p\right)}\mathrm{d}p}}
\left.+\frac{\sin\left(\frac{\pi }{2}\right)}{\pi }e^{-\frac{z^{2}}{2}}\int_{\eta }^{\infty}\frac{e^{-\frac{z^{2}x}{2}}}{\sqrt{x}\left(1+x\right)}\mathrm{d}x\right\} ,\label{eq:q-2}
\end{eqnarray}
with the contour of integration $C_{\eta }\left\backslash \left\{ -\eta \right\} \right.$ of
the first term being the counterclockwise running circle with a radius
$\eta<1$ excluding point $-\eta$.

Applying the polar substitution $p=\rho e^{i\varphi }$ in $\mathrm{Q}_{1}\left(\eta,z\right)$
and collapsing the contour $C_{\eta }$ yields:
\begin{eqnarray}
\underset{\eta \rightarrow0}{\lim}\mathrm{Q}_{1}\left(\eta,z\right)\!      \sim \! \left|\int_{-\pi }^{\pi }\frac{i\sqrt{\rho }e^{\frac{z^{2}}{2}\rho \cos\left(\varphi \right)}}{\left(1-\rho e^{i\varphi }\right)}e^{i\left(\frac{z^{2}}{2}\rho \sin\left(\varphi \right)+\varphi \right)}\mathrm{d}\varphi \;\right|_{\rho \rightarrow0}&\leq&
\int_{-\pi }^{\pi }\left|\frac{i\sqrt{\rho }e^{\frac{z^{2}}{2}\rho \cos\left(\varphi \right)}}{\left(1-\rho e^{i\varphi }\right)}e^{i\left(\frac{z^{2}}{2}\rho \sin\left(\varphi \right)+\varphi \right)}\right|_{\rho \rightarrow0}\mathrm{d}\varphi \; \nonumber \\
&\leq&
\int_{-\pi }^{\pi }\left.\frac{\sqrt{\rho }e^{\frac{z^{2}}{2}\rho \cos\left(\varphi \right)}}{\left(1-\rho \right)}\right|_{\rho \rightarrow0}\mathrm{d}\varphi \;               \rightarrow \!0.
\end{eqnarray}
Thus \eqref{eq:q-2} turns into:
\begin{equation*}
Q\left(z\right)=\frac{1}{2\pi }e^{-\frac{z^{2}}{2}}\int_{0}^{\infty}\frac{e^{-\frac{z^{2}x}{2}}}{\sqrt{x}\left(1+x\right)}\mathrm{d}x=\frac{1}{2}\mathrm{erfc}\left(\frac{z}{\sqrt{2}}\right).
\end{equation*}

The last equality relates Gauss Q-function with complementary error
function and is well known \cite{NIST}. Hence the contour-integral
representation in (\ref{eq:Lem1}) is justified.
\end{pf}
\begin{lem}
The integral of Gauss Q-function $Q\left(\sqrt{2\alpha \gamma }\right)$
weighted with the probability density function of Nakagami-m distribution
$f_{\gamma }\left(\gamma \right)$ can be represented as

\begin{equation}
\int_{0}^{\infty}Q\left(\sqrt{2\alpha \gamma }\right)f_{\gamma }\left(\gamma \right)d\gamma=\frac{1}{2}\mathrm{I}_{\frac{m}{m+\alpha \bar{\gamma }}}\left(m,\frac{1}{2}\right)\label{eq:Lem2}
\end{equation}
where $\mathrm{I}_{x}\left(y,z\right)$ is the regularized incomplete
beta function
\end{lem}
\begin{pf}
Denoting $\mathrm{J_{1}}={\displaystyle \int_{0}^{\infty}Q\left(\sqrt{2\alpha \gamma }\right)f_{\gamma }\left(\gamma \right)d\gamma }$
and applying Lemma~1 one obtains:
\begin{eqnarray}
\hspace{-5pt}\mathrm{J_{1}}&=&\frac{1}{2}\frac{1}{2\pi i}\int_{-\infty}^{\left(0_{+}\right)}\frac{1}{\sqrt{p}(1-p)}\int_{0}^{\infty}\mathit{f}_{\gamma }\left(\gamma \right)e^{\alpha (p-1)\gamma }\mathrm{d}\mathrm{\gamma d}p \nonumber\\
&=&\frac{1}{2}\frac{1}{2\pi i}\int_{-\infty}^{\left(0_{+}\right)}\frac{\mathcal{M}(\alpha (p-1))}{\sqrt{p}(1-p)}\mathrm{d}p,\label{eq:J1-1}
\end{eqnarray}
where $\mathcal{M}(                          \cdot )$ is the moment generating function (MGF),
defined as $\mathcal{M}(p)\triangleq\mathbb{E}\left\{ e^{p\gamma }\right\} $. Assuming that
for the Nakagami-m fading channel \cite{Sim05} MGF is given by $\mathcal{M}(p)=\left(1-\frac{p\bar{\gamma }}{m}\right)^{-m}$,
(\ref{eq:J1-1}) can be represented as:
\begin{equation}
\mathrm{J_{1}}{\displaystyle =\frac{b^{m}}{2\pi }\int_{0}^{\infty}\frac{\text{1}}{\sqrt{p}(1+p)(b+1+p)^{m}}\mathrm{d}p},\label{eq:J1-2}
\end{equation}
with $b=\frac{m}{\alpha \bar{\gamma }}$. The last equality was obtained
by deforming the integration contour as it was done in \cite{Gvo17}.

Using the integral representation of Gauss hypergeometric function
(see \S~2.2.6 in \cite{Pru92}) ${}_{2}F_{1}\left(a,b;c;z\right)$
yields:
\begin{eqnarray}
\mathrm{J}_{1}=\frac{b^{m}\mathrm{B}\left(\frac{1}{2},m+\frac{1}{2}\right)}{2\pi (1+b)^{2}}{}_{2}F_{1}\left(\frac{1}{2},m;m+1;\frac{b}{1+b}\right)
=\frac{b^{m}\mathrm{B}\left(\frac{1}{2},m+\frac{1}{2}\right)}{2\pi }\mathrm{B}_{\frac{b}{1+b}}\left(\frac{1}{2},m+\frac{1}{2}\right)
=\frac{1}{2}\mathrm{I}_{\frac{m}{m+\alpha \bar{\gamma }}}\left(m,\frac{1}{2}\right),\label{eq:J1-3}
\end{eqnarray}
where $\mathrm{I}_{x}\left(y,z\right)=\frac{\mathrm{B_{x}}\left(y,z\right)}{\mathrm{B}\left(y,z\right)}$
is the regularized incomplete beta function (see \S~8.17.2 in \cite{NIST}). Hence (\ref{eq:Lem2})
holds.
\end{pf}
\begin{lem}
The integral of squared Gauss Q-function $Q^{2}\left(\sqrt{2\alpha \gamma }\right)$
weighted with the probability density function of Nakagami-m distribution
$f_{\gamma }\left(\gamma \right)$ can be represented as
\begin{eqnarray}
&&\hspace{-10pt}\int_{0}^{\infty}Q^{2}\left(\sqrt{2\alpha \gamma }\right)f_{\gamma }\left(\gamma \right)d\gamma=\frac{\mathrm{I}_{\frac{m}{m+\alpha \bar{\gamma }}}\left(m,\frac{1}{2}\right)}{4}-R_{2}\left(\alpha,\bar{\gamma },m\right)\label{eq:Lem3}
\end{eqnarray}
where $R_{2}\left(\alpha,\bar{\gamma },m\right)$ is defined in ( \ref{eq:J2-8}).
\end{lem}
\begin{pf}
Denoting the initial integral as $\mathrm{J_{2}}={\displaystyle \int_{0}^{\infty}Q^{2}\left(\sqrt{2\alpha \gamma }\right)f_{\gamma }\left(\gamma \right)d\gamma }$
and sequentially applying Lemma~1 twice one obtains:
\begin{eqnarray}
\mathrm{J_{2}}=\frac{1}{2}\frac{1}{2\pi i}\int_{-\infty}^{\left(0_{+}\right)}\frac{\left\{ \frac{1}{2}\frac{1}{2\pi i}\int_{-\infty}^{\left(0_{+}\right)}\frac{\mathcal{M}(\alpha (q+p-2))}{\sqrt{q}(1-q)}\mathrm{d}q\right\} }{\sqrt{p}(1-p)}\mathrm{d}p\label{eq:J2-1}
\end{eqnarray}

The inner integral can be evaluated by applying the result of Lemma~2.
\begin{eqnarray}
\hspace{-12pt}\int_{-\infty}^{\left(0_{+}\right)}\frac{\mathcal{M}(\alpha (q+p-2))}{4\pi i\sqrt{q}(1-q)}\mathrm{d}q\!=\!\frac{b^{m}\left(1-\mathrm{I}_{\frac{1}{b+2-p}}\left(\frac{1}{2},m\right)\right)}{2(b+1-p)^{m}}\label{eq:J2-2}
\end{eqnarray}
Applying the property of the regularized incomplete beta function $\mathrm{I}_{x}\left(y,z\right)=1-\mathrm{I}_{1-x}\left(z,y\right)$
(see \S~8.17.4 in \cite{NIST}) gives:
\begin{eqnarray}
\hspace{-50pt}\mathrm{J_{2}}\!\!\!\!\!\!&=&\!\!\!\!\!\!\frac{1}{4\pi i}\frac{b^{m}}{2}\int_{-\infty}^{\left(0_{+}\right)}\frac{1}{\sqrt{p}(1-p)(b+1-p)^{m}}\mathrm{d}p-\underset{R_{2}\left(\alpha,\bar{\gamma },m\right)}{\underbrace{\frac{1}{4\pi i}\frac{b^{m}}{2}\int_{-\infty}^{\left(0_{+}\right)}\frac{\mathrm{I}_{\frac{1}{b+2-p}}\left(\frac{1}{2},m\right)}{\sqrt{p}(1-p)(b+1-p)^{m}}\mathrm{d}p}}\nonumber\\
\!\!\!\!\!\!&=&\!\!\!\!\!\!\frac{1}{4}\mathrm{I}_{\frac{m}{m+\alpha \bar{\gamma }}}\left(m,\frac{1}{2}\right)-R_{2}\left(\alpha,\bar{\gamma },m\right),\label{eq:J2-3}
\end{eqnarray}
where $R_{2}\left(\alpha,\bar{\gamma },m\right)$ is defined as
\begin{eqnarray}
\hspace{-8pt}R_{2}\left(\alpha,\bar{\gamma },m\right)\!=\!\frac{b^{m}}{4\pi }\int_{0}^{\infty}\!\frac{\mathrm{I}_{\frac{1}{b+2+p}}\left(\frac{1}{2},m\right)}{\sqrt{p}(1+p)(b+1+p)^{m}}\mathrm{d}p.\label{eq:J2-4}
\end{eqnarray}

Recall the series expansion of the regularized incomplete beta function:
(see \S~15.2.1 in \cite{NIST})
\begin{equation}
\mathrm{I}_{\frac{1}{b+2+p}}\left(\frac{1}{2},m\right)=\sum_{n=0}^{\infty}\frac{\left(1-m\right)_{n}(b+2+p)^{-n-\frac{1}{2}}}{\mathrm{B}\left(\frac{1}{2},m\right)n!\left(n+\frac{1}{2}\right)}.\label{eq:J2-5}
\end{equation}
Substituting (\ref{eq:J2-5}) into (\ref{eq:J2-2})
and performing
the change of integration variable $t=\frac{1}{1+p}$ yields:
\begin{eqnarray}
\hspace{-30pt}R_{2}\left(\alpha,\bar{\gamma },m\right)=\frac{b^{m}}{4\pi \mathrm{B}\left(\frac{1}{2},m\right)}\sum_{n=0}^{\infty}\frac{\left(1-m\right)_{n}}{n!\left(n+\frac{1}{2}\right)}                     \int_{0}^{1}\frac{\left(1+\left(1+b\right)t\right)^{-n-\frac{1}{2}}}{t^{-(m+n)}\left(1-t\right)^{\frac{1}{2}}\left(1+bt\right)^{m}}\mathrm{d}t.\label{eq:J2-7}
\end{eqnarray}

Using the definition of the Appell hypergeometric function $\mathrm{F}_{1}\left(\alpha,\beta,\beta';\gamma;x_{1};x_{2}\right)$
(see \S~16.15.1 \cite{NIST})
\begin{eqnarray}
\int_{0}^{a}x^{\alpha-1}\left(a-x\right)^{\beta-1}\left(1-ux\right)^{-\rho }\left(1-vx\right)^{-\lambda }\mathrm{d}x=a^{\alpha+\beta }\mathrm{B}\left(\alpha,\beta \right)\mathrm{F}_{1}\left(\alpha,\rho,\lambda;\alpha+\beta;ua;va\right),\label{eq:Appell}
\end{eqnarray}
the second term in (\ref{eq:J2-3}) can be represented as:
\begin{eqnarray}\label{eq:J2-8}
R_{2}\left(\alpha,\bar{\gamma },m\right)=\frac{\left(\frac{m}{\alpha \bar{\gamma }}\right)^{m}}{4\pi }\sum_{n=0}^{\infty}\frac{\left(1-m\right)_{n}\mathrm{B}\left(n+m+1,\frac{1}{2}\right)}{n!\left(n+\frac{1}{2}\right)\mathrm{B}\left(\frac{1}{2},m\right)}                     \mathrm{F}_{1}\left(n+m+1,m,n+\frac{1}{2};m+n+\frac{3}{2};-\frac{m}{\alpha \bar{\gamma }};-\left(1+\frac{m}{\alpha \bar{\gamma }}\right)\right).
\end{eqnarray}

Collecting (\ref{eq:J2-3}) and (\ref{eq:J2-8}) completes the proof.
\end{pf}
Although the derived expression is analytically rigorous, from a
practical viewpoint, the series (\ref{eq:J2-8}) has to be terminated,
but it should be noted that the series converges quite fast and only
a few terms are needed to obtain high accuracy (see Section \ref{sec:Sim}).
Moreover, even though the derived solution is given in terms of the
hypergeometric function of two variables, one can remark that it is
readily accessible in modern computer-algebra systems, like Wolfram
MATHEMATICA, Matlab, MAPLE, etc., delivering high computational efficiency
of (\ref{eq:Lem3}).

The obtained expressions help to formulate the final result.
\begin{thm}
For Nakagami-m multipath fading channel the closed-form and approximating expressions for the average bit error rate
of M-QAM are given by
\begin{equation}
\overline{\mathrm{BER}}_{th}\!=\!\int_{0}^{\infty}\mathrm{BER}_{th}\left(\gamma \right)f_{\gamma }\left(\gamma \right)d\gamma \!=\!c_0\mathrm{I}_{\frac{m}{m+c_1\bar{\gamma }}}\left(m,\frac{1}{2}\right)+4c_0^2R_{2}\left(c_1,\bar{\gamma },m\right),\label{eq:ABERth}
\end{equation}
\begin{eqnarray}
\hspace{-20pt}\overline{\mathrm{BER}}_{ap}=\int_{0}^{\infty}\mathrm{BER}_{ap}\left(\gamma \right)f_{\gamma }\left(\gamma \right)d\gamma =2c_0\sum_{j=1}^{\sqrt{M}/2}\mathrm{I}_{\frac{m}{m+c_1(2j-1)^2\bar{\gamma }}}\left(m,\frac{1}{2}\right),\label{eq:ABERap}
\end{eqnarray}
with coefficients $c_0, c_1$ being defined as in \eqref{eq:BERth}.
\end{thm}
\begin{pf}
To prove the first part of Theorem 4, one should substitute the closed-form BER definition \eqref{eq:ABERth} and apply the  results of Lemma 2 and Lemma 3, i.e.,
\begin{eqnarray}
\hspace{-20pt}\overline{\mathrm{BER}}_{th}=\int_{0}^{\infty}\mathrm{BER}_{th}\left(\gamma \right)f_{\gamma }\left(\gamma \right)d\gamma \!\!\!\!\!\!\!&=&\!\!\!\!\!\!\!
4c_{0}\int_{0}^{\infty}Q\left(\sqrt{2c_{1}\gamma }\right)f_{\gamma }\left(\gamma \right)d\gamma-
4c_{0}^{2}\int_{0}^{\infty}Q^{2}\left(\sqrt{2c_{1}\gamma }\right)f_{\gamma }\left(\gamma \right)d\gamma  \nonumber \\ &=&\!\!\!\!\!\!\!c_0\mathrm{I}_{\frac{m}{m+c_1\bar{\gamma }}}\left(m,\frac{1}{2}\right)+4c_0^2R_{2}\left(c_1,\bar{\gamma },m\right),
\end{eqnarray}

The second part of Theorem 4 can be justified in a similar manner: one should combine BER approximation \eqref{eq:ABERap} and utilize the expression derived in Lemma 2, i.e.,
\begin{eqnarray}
\overline{\mathrm{BER}}_{ap}=\int_{0}^{\infty}\mathrm{BER}_{ap}\left(\gamma \right)f_{\gamma }\left(\gamma \right)d\gamma \!\!\!\!\!\!\!&=&\!\!\!\!\!\!\!
4c_0\sum_{j=1}^{\sqrt{M}/2}\int_{0}^{\infty}Q\left(\sqrt{c_1(2j-1)^2\gamma }\right)f_{\gamma }\left(\gamma \right)d\gamma \nonumber\\
\!\!\!\!\!\!\!&=&\!\!\!\!\!\!\!2c_0\sum_{j=1}^{\sqrt{M}/2}\mathrm{I}_{\frac{m}{m+c_1(2j-1)^2\bar{\gamma }}}\left(m,\frac{1}{2}\right)
\end{eqnarray}
\end{pf}

\begin{figure}[!b]
\centerline{\includegraphics[width=\columnwidth]{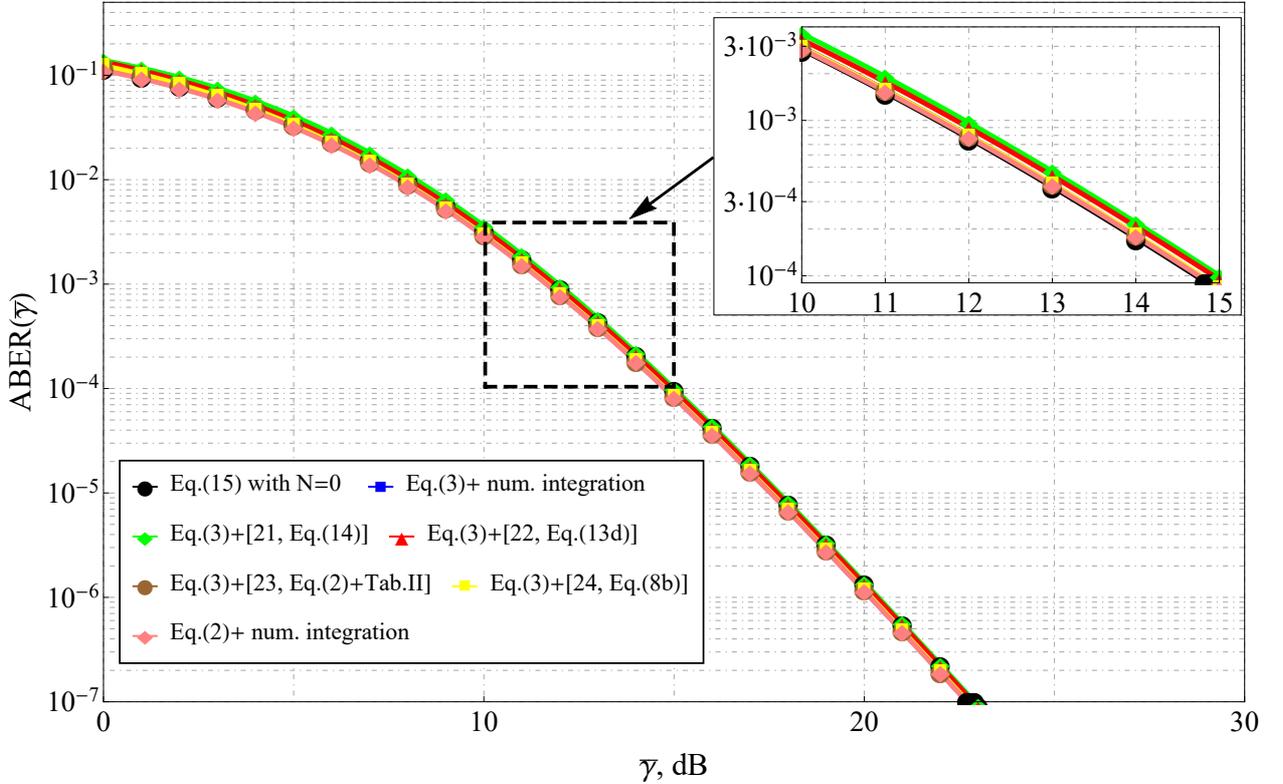}}
\caption{Average BER for light fading and low-order modulation (i.e., $m=4.1$ and $8$-QAM)}
\label{fig1}
\end{figure}

From the practical perspective, the evaluation of \eqref{eq:ABERth} needs infinite summation in \eqref{eq:J2-8}, but as it will be demonstrated in Section IV, it can be efficiently truncated to only few first terms. Thus, for further calculations \eqref{eq:J2-8} will be replaced with its substitute a truncated to $N$-term version of, i.e. $R_{2}\left(N,c_1,\bar{\gamma },m\right)$.

To this extent, one can see that due to the intimate connections between incomplete Gamma and Beta functions with Gauss hypergeometric function (see \eqref{eq:J1-3}), \eqref{eq:Lem2}, although being a compact and neat new notation, can be rearranged in a well-known form (see \cite{Sim05}). On the other hand, the representations \eqref{eq:J1-1}, \eqref{eq:J2-1} and the proposed procedure of evaluating the initial integrals \eqref{eq:Lem2} and \eqref{eq:Lem3} are novel and can be further extended to various channel models.

It should be specifically pointed out that the derived closed-form solution \eqref{eq:J2-8} lacks the drawback of validity under the assumption of an integer $m$, inherited by the existing solution (see, \cite{Sim05} Section $5.1.6$, equation $(5.30)$), and thus is valid for arbitrary values of fading parameter. Therefore, for further numerical evaluations, only non-integer $m$ will be assumed (i.e. that are not covered by the existing closed-form results).

\begin{figure}[!t]
\centerline{\includegraphics[width=\columnwidth]{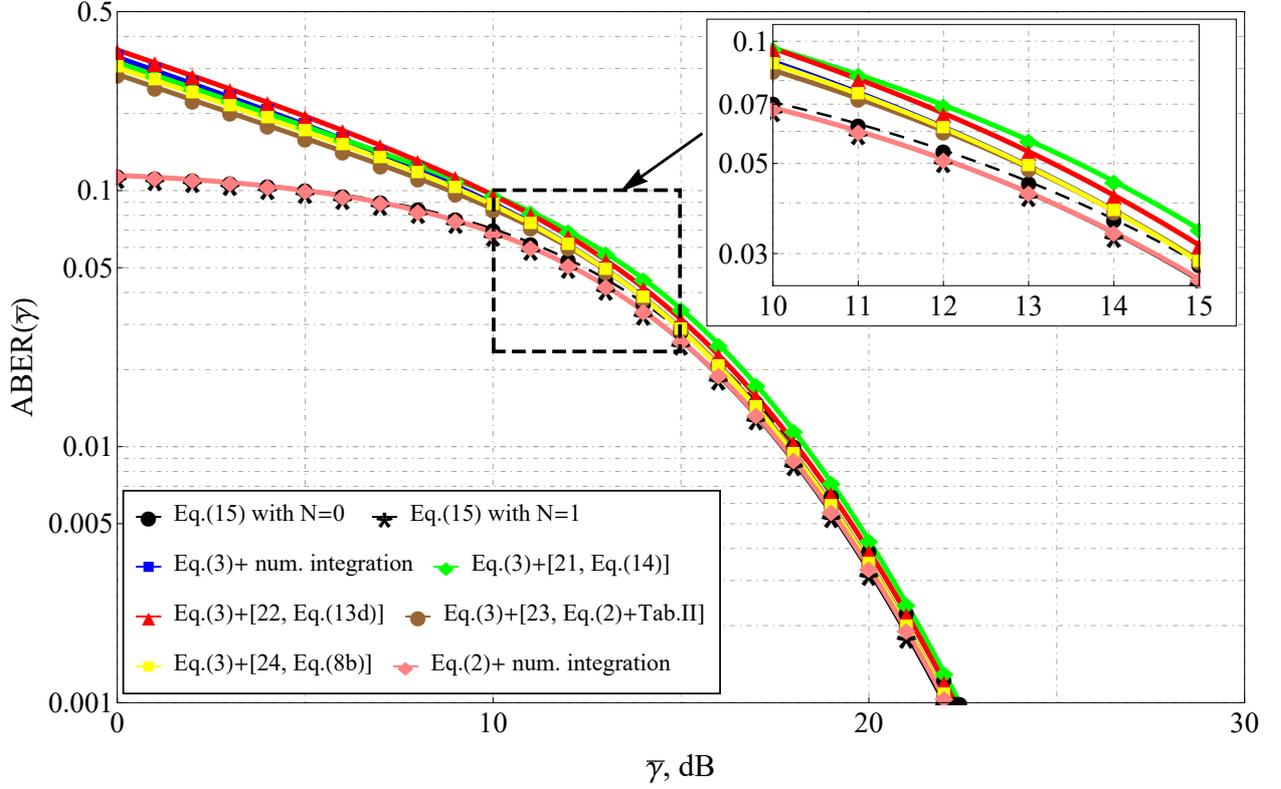}}
\caption{Average BER for light fading and high-order modulation (i.e., $m=4.1$ and $256$-QAM)}
\label{fig2}
\end{figure}
\section{\label{sec:Sim}Numeric simulation and analysis}
\subsection{ABER analysis}
To verify the correctness of the proposed results, the numeric analysis of the average BER was performed for high and low modulation orders combined with heavy and light fading (see Fig~(\ref{fig1})-Fig~(\ref{fig4}) respectively). The derived solution \eqref{eq:ABERth} was compared with the brute-force numeric integration of \eqref{eq:ABER} with: closed-form BER expression \eqref{eq:BERth} (plotted with pink lines and pink diamond-shaped markers) and with approximated BER expression \eqref{eq:ABERap} (plotted with blue lines and blue square markers). The obtained solution \eqref{eq:ABERth} for all of the assumed situations was restricted only to the one-term series truncation ($N=0$) (solid black line with black circle markers). In addition to that, a two-term ($N=1$) truncation curve was plotted in Fig~(\ref{fig4}) (depicted with a solid black line and black star markers)) to demonstrate approximation quality improvement. Those analytic results were compared with the solutions derived under Gaussian Q-function exponential-type approximation (\cite{Chi03, Los09, Ola12, Sad17}), which differ in the way the approximating coefficients are defined: in \cite{Chi03} they are found by minimizing the integral of the relative error in the range of values of interest, in \cite{Sad17} they are derived via trapezoidal-rule integration with $4$ subintervals (defined as optimal), in \cite{Los09} a three-term Prony series approximation is used and in \cite{Ola12} obtained by minimizing the log-scaled relative minimum mean-squared error.

\begin{figure}[!t]
\centerline{\includegraphics[width=\columnwidth]{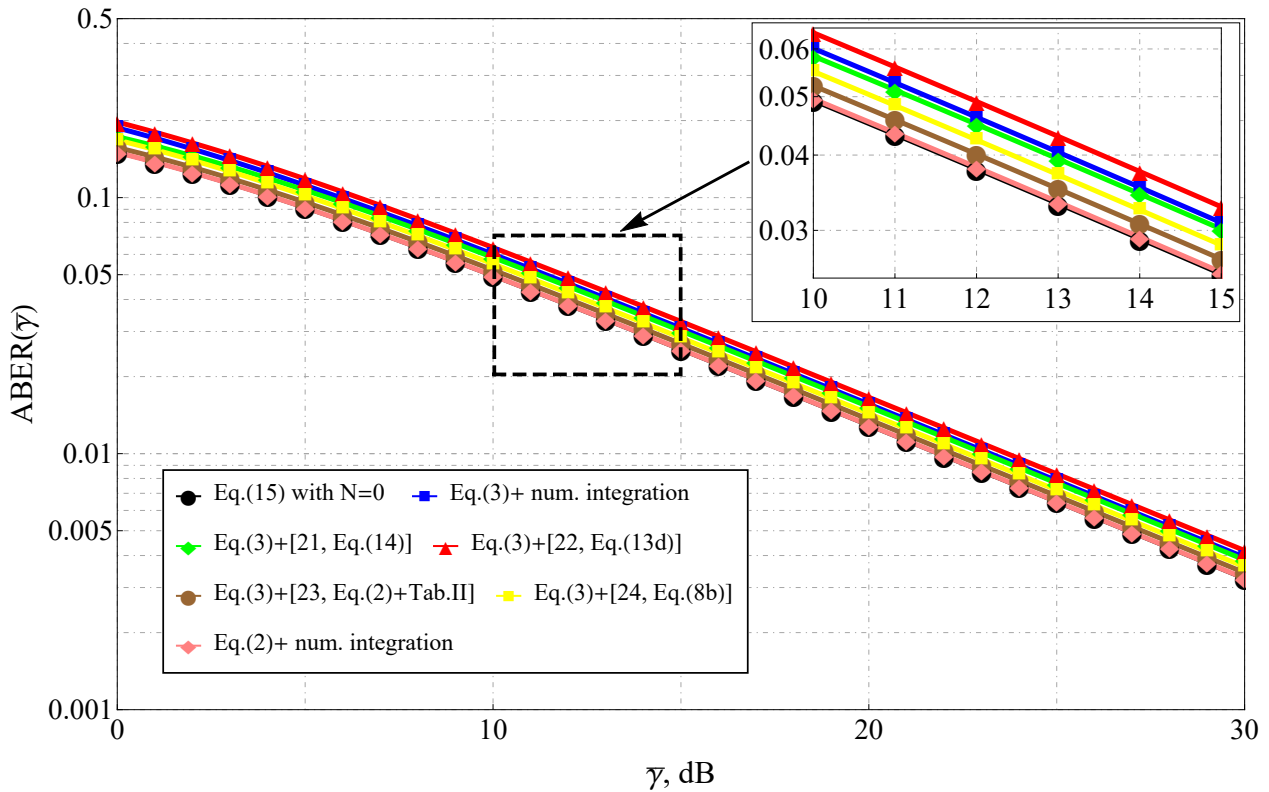}}
\caption{Average BER for heavy fading and low-order modulation (i.e., $m=0.6$ and $8$-QAM)}
\label{fig3}
\end{figure}

\begin{figure}[!t]
\centerline{\includegraphics[width=\columnwidth]{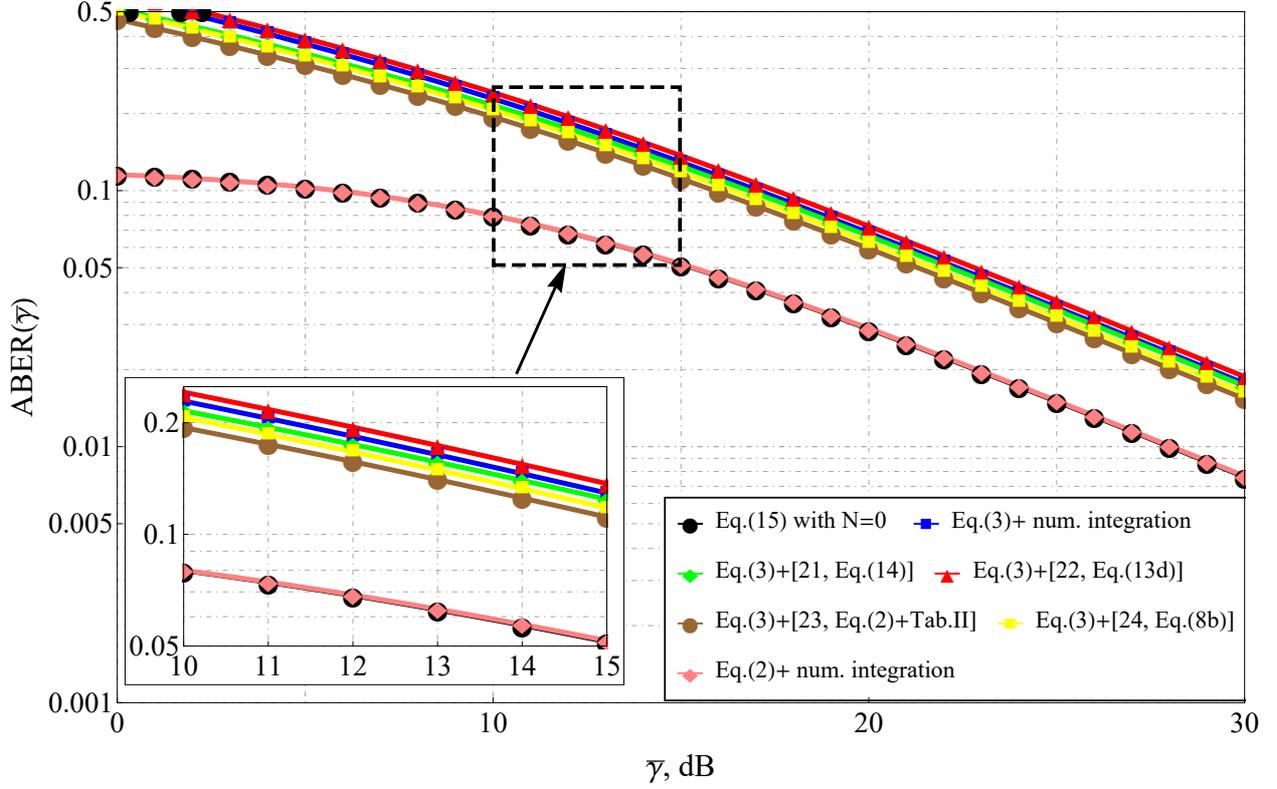}}
\caption{Average BER for heavy fading and high-order modulation (i.e., $m=0.6$ and $256$-QAM)}
\label{fig4}
\end{figure}

The performed analysis demonstrated that all of the prevailing approximations work fine for light shadowing and low-order modulation schemes (see Fig~(\ref{fig1} and Fig~(\ref{fig3})), but degrading the propagation conditions (or increasing modulation) (see Fig~(\ref{fig2})and Fig~(\ref{fig4})) heavily impairs approximation quality: the energy efficiency loss is from 0.5~dB up to 1~dB for $m=4.1$ (depending on the error level) and from 1~dB to 15~dB for $m=0.6$. At the same time, the proposed solution (even in the case of the single-term series truncation) delivers excellent results for all of the assumed scenarios.

\subsection{Proposed approximation analysis}
\begin{figure}[!t]
\centerline{\includegraphics[width=\columnwidth]{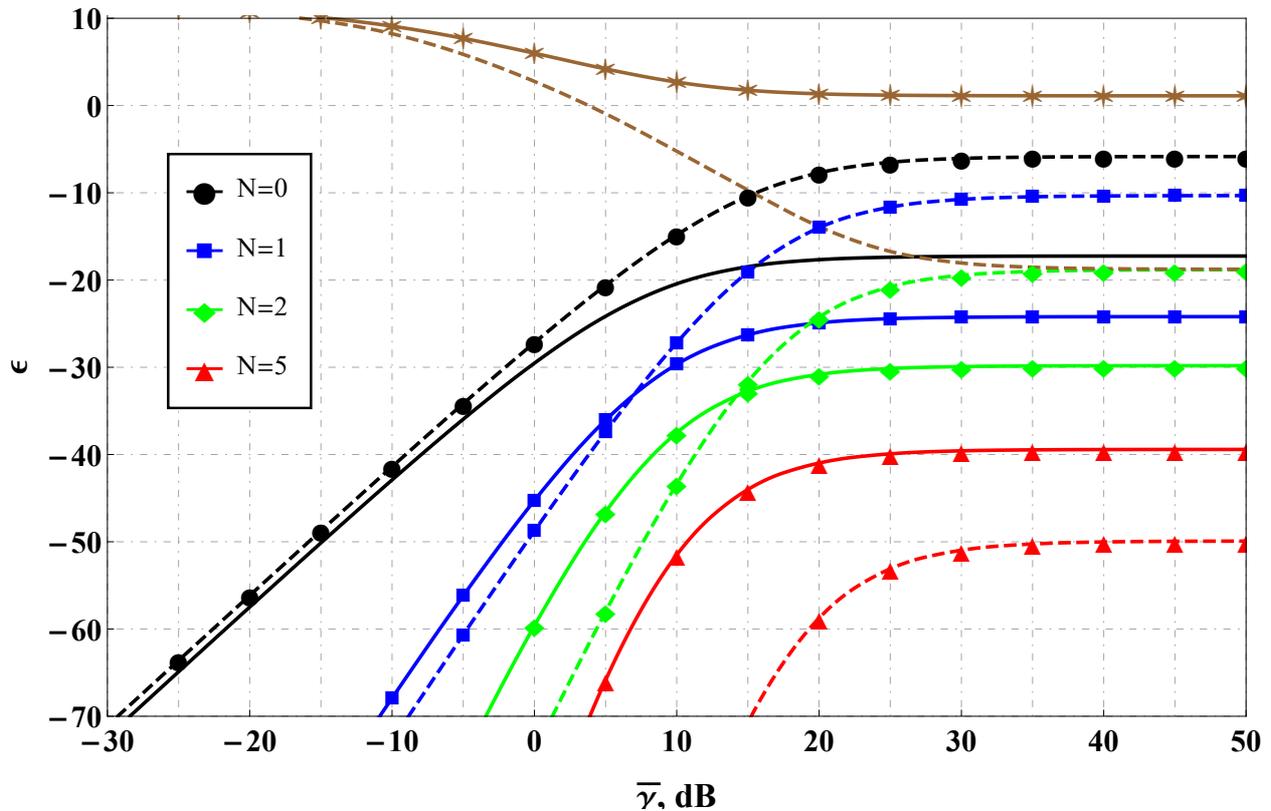}}
\caption{Logarithmically-scaled discrepancy $\epsilon$ for different number of summands in (\ref{eq:J2-8})}
\label{fig5}
\end{figure}

As it was already mentioned, the number of summands in (\ref{eq:ABERth}) that deliver the desired quality (ABER) depends on the average SNR, and even the first few terms that can provide reasonable quality. To estimate the loss of accuracy
(see Fig~(\ref{fig5})) one assumes the logarithmically-scaled discrepancy
$\epsilon=10\lg\left|\frac{\overline{\mathrm{BER}}_{num}\left(\overline{\gamma }\right)-\overline{\mathrm{BER}}\left(\overline{\gamma }\right)}{\overline{\mathrm{BER}}_{num}\left(\overline{\gamma }\right)}\right|$
between $\overline{\mathrm{BER}}_{num}\left(\overline{\gamma }\right)$,
that is computed via numerical integration of (\ref{eq:ABER}) (with
(\ref{eq:BERth}) being substituted) and  $\overline{\mathrm{BER}}\left(\overline{\gamma }\right)$ equals to
\begin{itemize}
  \item $\overline{\mathrm{BER}}_{th}\left(\overline{\gamma }\right)$  given by the derived closed-form solution (\ref{eq:ABERth})truncated to $N$ terms (black, blue, green and red lines);
  \item $\overline{\mathrm{BER}}_{ap}\left(\overline{\gamma }\right)$ given by the derived approximation (\ref{eq:ABERap}) (brown lines);
\end{itemize}

\begin{figure}[!t]
\centerline{\includegraphics[width=\columnwidth]{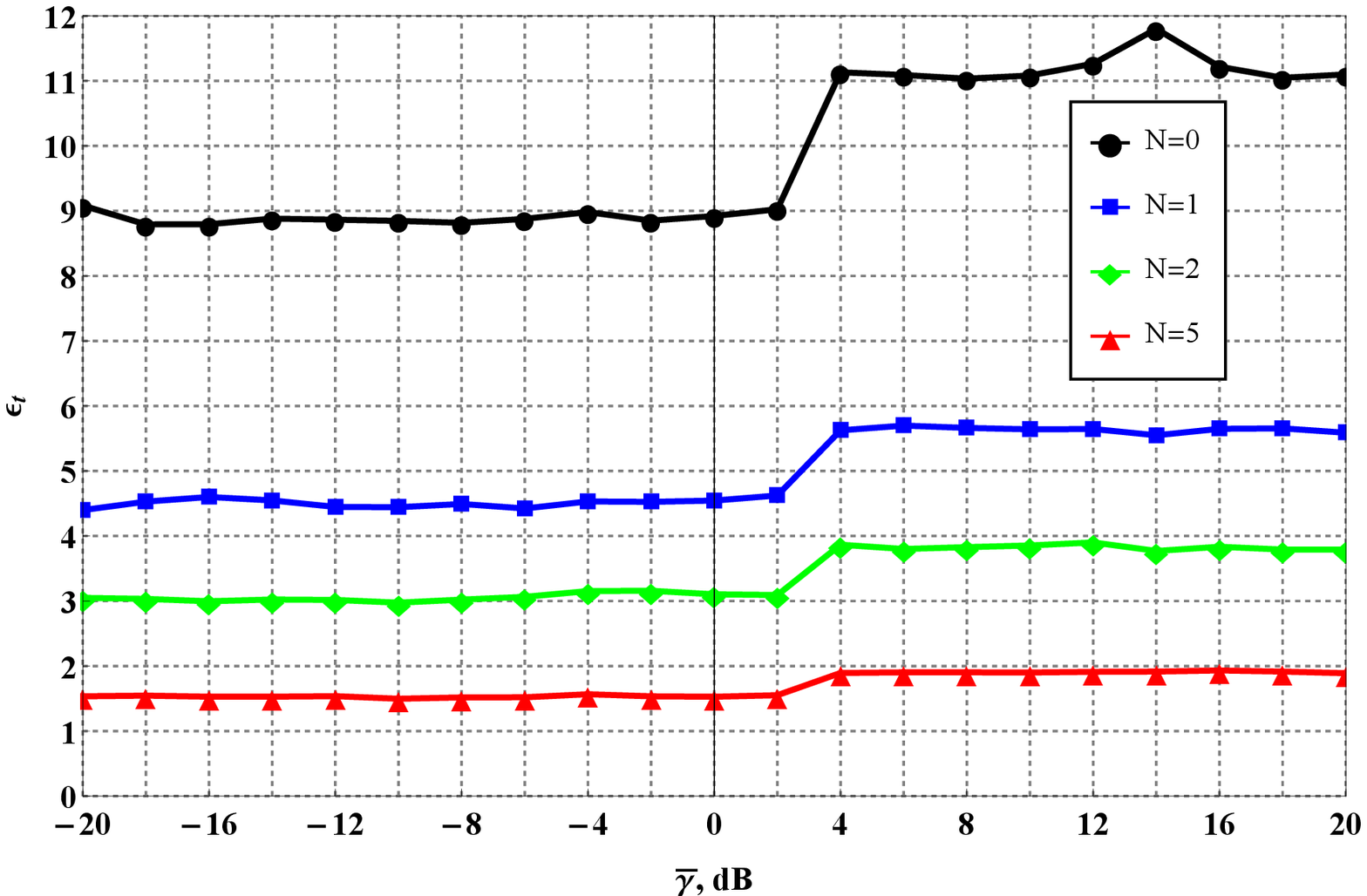}}
\caption{Computational time gain $\epsilon_t$ for different number of summands in (\ref{eq:J2-8})}
\label{fig6}
\end{figure}

The results were obtained under heavy $m=0.6$ (solid lines) and
light fading $m=4.1$ (dashed lines) for 256-QAM. It is clear that for heavy fading (which is of primary interest), even the zero-term truncation in (\ref{eq:ABERth}) yields better performance compared to the existing approximation (see \cite{Lu99}); for light fading one needs more than a single term, but still, even three terms are enough. Moreover, increasing $N$ up to two to five terms improves the accuracy by several
orders of magnitude.  It can be noticed that even in the case of not very high $\overline{\gamma }$ (up to 15~dB), the approximating expression is strictly worse (in terms of $\epsilon$) than the proposed solution; this is mainly due to the fact that \eqref{eq:ABERap} is commonly considered for the case of large symbol SNR, for which the only significant symbol errors are those that occur in adjacent signal levels (see \cite{Sim05}, section~5.1.4).

It can be argued whether the computational accuracy is achieved at
the expense of computational time. Therefore, computational time gain defined as $\epsilon_{t}=\frac{\Delta t_{th}}{\Delta t_{num}}$ was analyzed (with $\Delta t_{th}$ is the interval time needed to compute $\overline{\mathrm{BER}}_{th}\left(\overline{\gamma }\right)$ defined as earlier, and $\Delta t_{num}$ is the same for $\overline{\mathrm{BER}}_{num}\left(\overline{\gamma }\right)$ with the same five digit precision). The performed numerical simulation of the direct comparison between the two (see (\ref{fig6}) for the case of $m=0.6$ and 256-QAM) demonstrated that $\epsilon_{t}\geq 1$ for the proposed expression, delivering computational speedup from 1.5 to 12 times depending
on the number of summands (from 0 to 5).

\section{Further generalization and discussion}

It can be pointed out that the solutions to a large number of modern wireless communications problems depend heavily on the normality assumption of some stochastic physical effects (e.g., reflection, multiple diffractions, fading); thus, the expressions with Gauss Q-function and its squared version naturally arise. An important example is the problem of signal detection with the help of the energy-based spectrum sensing procedure for ad hoc systems and networks, including polar-coding transmission for 5G \cite{Ju21}, MIMO-OFDM Cognitive Radio Systems \cite{Lor22}, listen-before-talk procedure in 5G networks \cite{Nik21}, etc.

Since the proposed research makes use of an MGF approach, and a wide range of channels (including non-line-of-sight and shadowed line-of-sight) have the same power-type MGFs (including $\eta-\mu$ \cite{Pep11}, Hoyt \cite{Pep11}, complex Nakagami-m \cite{Yac05}, etc.), the derived results can be further generalized for those models. Nevertheless, the possible overcomplication of the derived mathematical description for such generalizations and its applicability in engineering practice is still an open question.

Other possible extensions of the proposed research encompass wireless physical layer security and higher-order capacity analysis due to the existing connection between the MGF and the spectral efficiency \cite{Yil12}.

\section{Conclusions}

Closed-form average bit error rate calculation for wireless communication systems in the presence multipath fading is an important tool that helps to solve various engineering problems. At the same time, it requires the solutions of integrals involving the Gauss Q-function and its square, which for many fading channel models cannot be obtained. One of the possible solutions relies upon the widely renowned moment generating function approach. Its generalization that makes use of contour-integral representation of several more general special functions can be successfully applied to yield analytic solutions. This research has presented a novel procedure for evaluating such integrals for the case of the Nakagami-m fading channel. A new notation for the closed-form ABER of the QAM signal and its approximate version were evaluated.  The
correctness of the obtained solution was verified, and its computational
efficiency (in terms of accuracy and time gain), compared to the
prevailing approximation was demonstrated.

%


\end{document}